\documentclass[aps,prb,preprint,groupedaddress,showpacs]{revtex4}

\usepackage{graphicx}
\usepackage{dcolumn}
\usepackage{bm}
\usepackage{subfigure}

\begin{document}


\title{Finite-Size Scaling of the Domain Wall Entropy Distributions
for the 2D $\pm J$ Ising Spin Glass}


\author{Ronald Fisch}
\email[]{ron@princeton.edu}
\affiliation{382 Willowbrook Dr.\\
North Brunswick, NJ 08902}


\date{\today}

\begin{abstract}
The statistics of domain walls for ground states of the 2D Ising
spin glass with +1 and -1 bonds are studied for $L \times L$ square
lattices with $L \le 48$, and $p$ = 0.5, where $p$ is the fraction
of negative bonds, using periodic and/or antiperiodic boundary
conditions. When $L$ is even, almost all domain walls have energy
$E_{dw}$ = 0 or 4.  When $L$ is odd, most domain walls have $E_{dw}$
= 2.  The probability distribution of the entropy, $S_{dw}$, is
found to depend strongly on $E_{dw}$.  When $E_{dw} = 0$, the
probability distribution of $|S_{dw}|$ is approximately exponential.
The variance of this distribution is proportional to $L$, in
agreement with the results of Saul and Kardar.   For $E_{dw} = k~>~
0$ the distribution of $S_{dw}$ is not symmetric about zero.  In
these cases the variance still appears to be linear in $L$, but the
average of $S_{dw}$ grows faster than $\sqrt{ L }$.  This suggests a
one-parameter scaling form for the $L$-dependence of the
distributions of $S_{dw}$ for $k > 0$.

\end{abstract}

\pacs{75.10.Nr, 75.40.Mg, 75.60.Ch, 05.50.+q}

\maketitle

\section{INTRODUCTION}

There continues to be a controversy about the nature of the Ising
spin glass.  The Sherrington-Kirkpatrick model,\cite{SK75} with its
infinite-range interactions between the spins, is described by the
Parisi replica-symmetry breaking mean-field
theory.\cite{Par80,Par83} To understand models with short-range
interactions on finite-dimensional lattices, however, it is
necessary to include the effects of interfaces, which do not exist
in a well-defined way in an infinite-range model.  The droplet model
of Fisher and Huse,\cite{FH86,HF86,FH88} which starts from the
domain-wall renormalization group ideas of
McMillan\cite{McM84a,McM84b,McM85} and Bray and Moore,\cite{BM85,
BM86,BM87} and studies the properties of interfaces, provides a very
different viewpoint on the spin-glass phase.

In two dimensions (2D), the spin-glass phase is not stable at finite
temperature.  Because of this, it is necessary to treat cases with
continuous distributions of energies (CDE) and cases with quantized
distributions of energies (QDE) separately.\cite{BM86,AMMP03}

In three or more space dimensions, where a spin-glass phase is
believed to occur at finite temperature $T$, the general framework
of thermodynamics requires that the CDE and the QDE should be
treated on the same footing. The way this comes about is that in
these cases the typical domain wall energy increases as a positive
power of the size of the lattice. Thus the quantization energy
becomes a negligible fraction of the domain wall energy for large
lattices.  All bond distributions behave in a qualitatively similar
way, except that the QDE have finite ground state
entropies.\cite{BM86,FH88}

Amoruso, Hartmann, Hastings and Moore\cite{AHHM06} have recently
proposed that in 2D there is a relation
\begin{equation}
  d_S = 1 + {3 \over {4 ( 3 + \theta_E )}}   \, ,
\end{equation}
where $d_S$ is the fractal dimension of domain walls, and $\theta_E$
is the exponent which characterizes the scaling of the domain wall
energy with size.  For the CDE case, the existing numerical
estimates of $d_S$ and $\theta_E$ satisfy Eqn.~(1).  However, it is
unclear if Eqn.~(1) should continue to be correct when the scaling
exponent for spin correlations, $\eta$, is not zero. For the QDE,
the current estimates\cite{KL05,PB05} find $\eta \approx 0.14$.

In three dimensions it is known from the droplet
theory,\cite{HF86,BM87,FH88} that for the QDE, which have a positive
entropy at $T = 0$, in the spin-glass phase
\begin{equation}
  d_S = 2 \theta_S   \, .
\end{equation}
$\theta_S$ is the exponent for the scaling of domain wall entropy
with size.  Thus, for the QDE, Eqn.~(1) provides a relation between
the scaling of the energy and the entropy of domain walls. It is not
known how to calculate $d_S$ directly for the QDE case, so we need
to use Eqn.~(2) to check Eqn.~(1) in that case.  One might hope that
this relation would also hold in 2D, even though the spin-glass
order only occurs at $T = 0$.

For the QDE, it is known that $\theta_E = 0$.\cite{AMMP03,HY01} Then
using Eqn.~(1) gives $d_S = 5/4$, or using Eqn. (2), $\theta_S =
5/8$. The calculation of $\theta_S$ by Saul and
Kardar,\cite{SK93,SK94} found $\theta_S = 0.49 \pm 0.02$.  Since
$d_S$ cannot be less than 1, this result was interpreted as a strong
indication that $\theta _S = 1/2$.

In this work we will find that Eqn.~(1) may not work for the QDE
case in 2D.  It appears, however, that Eqn. (2) is still correct in
2D, except when the domain wall energy, $E_{dw}$, is zero.  The
actual behavior of the QDE probability distributions under
finite-size scaling turns out to be more subtle than what has been
assumed until recently.\cite{Fis06,Fis05} As pointed out by Wang,
Harrington and Preskill,\cite{WHP03} domain walls of zero energy
which cross the entire sample play a special role when the energy is
quantized.

We will analyze data for the $E_{dw}$ and for the domain wall
entropy, $S_{dw}$, for the ground states (GS) of 2D Ising spin
glasses obtained using a slightly modified version of the computer
program of Gallucio, Loebl and Vondr\'{a}k,\cite{GLV00} which is
based on the Pfaffian method.  The Pfaffians are calculated using a
fast exact integer arithmetic procedure, coded in C++.  Thus, there
is no roundoff error in the calculation until the double precision
logarithm is taken to obtain $S_{dw}$.  This extended precision is
essential, in order to obtain meaningful results for entropy
differences at large $L$.  An earlier version of the domain wall
entropy calculation,\cite{Fis05} using data provided by S. N.
Coppersmith,\cite{LC01} was limited to small $L \times L$ lattices
with even $L$ and came to somewhat different conclusions.

We will demonstrate that for $L \times L$ square lattices the
Edwards-Anderson\cite{EA75} (EA) model with a $\pm J$ bond
distribution has a strong correlation between $E_{dw}$ and $S_{dw}$
for the GS domain walls.  Because of this correlation, we will need
to treat domain walls of different energies as distinct classes.  We
will find that the scaling parameter identified by Saul and
Kardar\cite{SK93,SK94} is the one associated with domain walls
having $E_{dw} = 0$.  It is not, however, the one which controls the
dominant behavior for large $L$.

The Hamiltonian of the EA model for Ising spins is
\begin{equation}
  H = - \sum_{\langle ij \rangle} J_{ij} \sigma_{i} \sigma_{j}   \, ,
\end{equation}
where each spin $\sigma_{i}$ is a dynamical variable which has two
allowed states, +1 and -1.  The $\langle ij \rangle$ indicates a sum
over nearest neighbors on a simple square lattice of size $L \times
L$. We choose each bond $J_{ij}$ to be an independent identically
distributed quenched random variable, with the probability
distribution
\begin{equation}
  P ( J_{ij} ) = p \delta (J_{ij} + 1)~+~(1 - p) \delta (J_{ij} -
  1)   \, ,
\end{equation}
so that we actually set $J = 1$, as usual.  Thus $p$ is the
concentration of antiferromagnetic bonds, and $( 1 - p )$ is the
concentration of ferromagnetic bonds.

\section{GROUND STATE DOMAIN WALLS}

We define the GS entropy to be the natural logarithm of the number
of ground states.  For each sample the GS energy and GS entropy were
calculated for the four combinations of periodic (P) and
antiperiodic (A) toroidal boundary conditions along each of the two
axes of the square lattice.  We will refer to these as PP, PA, AP
and AA.  In the spin-glass region of the phase diagram, the
variation of the sample properties for changes of the boundary
conditions is small compared to the variation between different
samples of the same size,\cite{SK94} except when $p$ is close to the
ferromagnetic phase boundary and the ferromagnetic correlation
length becomes comparable to $L$.

We define domain walls for the spin glass as it was done in the
seminal work of McMillan.\cite{McM84b}  We look at differences
between two samples with the same set of bonds, and the same
boundary conditions in one direction, but different boundary
conditions in the other direction.  Thus, for each set of bonds we
obtain domain wall data from the four pairs (PP,PA), (PP,AP),
(AA,PA) and (AA,AP).  The reader should remember that the term
``domain wall", as used in this work, refers only to this
procedure. Saul and Kardar\cite{SK93,SK94} follow the same
procedure used in this work, but use the term ``defect" instead of
``domain wall".

The domain-wall renormalization group of McMillan\cite{McM84a} is
based on the idea that we are studying an effective coupling
constant which is changing with $L$.  For the CDE case we can use
the energy as the coupling constant. For the quantized energy case,
what we need to do is a slight generalization of this idea. We
should think of the coupling constant as the free energy at some
infinitesimal temperature.  When we do this, the entropy contributes
to the coupling constant.  As we will see, the distribution of
$E_{dw}$ rapidly becomes essentially independent of $L$ as $L$
becomes large, except that there are separate distributions for even
$L$ and odd $L$.  Under these conditions, it becomes possible to
treat each value of $E_{dw}$ as a separate class, representing a
different coupling constant.

The domain wall entropy, $S_{dw}$, is defined, by analogy to
$E_{dw}$, to be the difference in the GS entropy when the boundary
condition is changed along one direction from P to A (or vice
versa), with the boundary condition in the other direction remaining
fixed.  $[S_{dw}]$, where the brackets [~] indicate an average over
random samples of the $J_{ij}$, is expected to increase as a
positive power of $L$ for any $E_{dw}$.  Therefore, these coupling
constants must eventually, at large enough $L$, be controlled by
$[S_{dw}]$ for any $T > 0$. Of course, the value of $L$ which is
needed for this to happen depends in $T$. The droplet model assumes
that all these coupling constants, except for the $E_{dw} = 0$ case
which has a special symmetry, are equal.

As long as $E_{dw} > 0$, the two boundary conditions which we are
comparing are not on an equal footing.  As Wang, Harrington and
Preskill\cite{WHP03} express the situation, the $E_{dw} > 0$ domain
wall does not destroy the topological long-range order.  However, in
the $E_{dw} = 0$ case the two boundary conditions are on an equal
footing, and the topological order is destroyed.  Therefore the
$E_{dw} = 0$ class of domain walls can be expected to behave in a
special way, which differs from the prediction of the droplet model.

It is natural to wonder if topological long-range order can be
related to replica-symmetry breaking, and if the $E_{dw} = 0$ domain
walls can be described by the replica-symmetry breaking theory.  We
will not attempt to do this here.

It is important to realize that the meaning of a domain wall is very
different when the GS entropy is positive, as in the model we study
here, as compared to the standard case of a doubly degenerate ground
state.  In the standard case one can identify a line of bonds which
forms a boundary between regions of spins belonging to the two
different ground states.  It is not possible, in general, to do that
when there are many ground states.  Despite this, we continue to use
the term ``domain wall".

When $L$ is even, the energy difference, $E_{dw}$, for any pair must
be a multiple of 4.  When $L$ is odd, $E_{dw}$ is $4 n + 2$, where
$n$ is an integer.\cite{BM86}  The sign of $E_{dw}$ for a pair is
essentially arbitrary for $p = 1/2$. Thus we can, without loss of
generality, choose all of the domain-wall energies to be
non-negative.

\section{NUMERICAL RESULTS}

Our calculated statistics for $E_{dw}$ at $p = 0.5$, as a function
of $L$, for even $L$ and odd $L$ are given in Table~I and Table~II,
respectively.  For each $L$, 500 distinct random configurations of
bonds were studied.  We obtain four McMillan pairs for each random
sample, so we have 2000 sets of $E_{dw}$ and $S_{dw}$ at each $L$.
For even $L > 10$ it turns out, crudely speaking, that about 77\% of
the time we find $E_{dw} = 0$, and 23\% of the time $E_{dw} = 4$.
For odd $L > 20$, $E_{dw} = 2$ about 98.5\% of the time.  No domain
walls with energies greater than 8 were observed at any $L$ for
these values of $p$. This, however, does not have much fundamental
significance. The probability distribution for $E_{dw}$ is also a
weak function of $p$,\cite{Fis05} and a strong function of the
aspect ratio of the lattice.\cite{HBCMY02} Our results are
consistent with the results of Amoruso {\it et al.}\cite{AMMP03}

\begin{table}
\caption{\label{tab:table1}Domain wall energy statistics for $p =
0.5$ with even $L$.  The number of random bond configurations
studied for each $L$ was 500, and there are four McMillan pairs for
each of these.  $n_i$ is the number of domain walls of each type
having $E_{dw} = k$.  $f_k$ is the fraction of domain walls having
$E_{dw} = k$.}
\begin{ruledtabular}
\begin{tabular}{rrrrll}
 $L$ & $n_0$ & $n_4$ & $n_8$ & $f_0$ & $f_4$ \\
\hline
  8 & 1467 & 530 & 3 & 0.7335& 0.265 \\
 12 & 1542 & 458 & 0 & 0.771 & 0.229 \\
 16 & 1515 & 484 & 1 & 0.7575& 0.242 \\
 24 & 1578 & 422 & 0 & 0.789 & 0.211 \\
 32 & 1530 & 470 & 0 & 0.765 & 0.235 \\
 48 & 1546 & 450 & 4 & 0.773 & 0.225 \\

\end{tabular}
\end{ruledtabular}
\end{table}

\begin{table}
\caption{\label{tab:table2}Domain wall energy statistics for $p =
0.5$ with odd $L$.  Column labels as in Table~I.}
\begin{ruledtabular}
\begin{tabular}{rrrl}
 $L$ & $n_2$ & $n_6$ & $f_2$ \\
\hline
  7 & 1944 & 56 & 0.972 \\
 11 & 1960 & 40 & 0.980 \\
 15 & 1957 & 43 & 0.9785\\
 21 & 1973 & 27 & 0.9865\\
 29 & 1967 & 33 & 0.9835\\
 41 & 1973 & 27 & 0.9865\\
\end{tabular}
\end{ruledtabular}
\end{table}

It is interesting to note that Wang, Harrington and
Preskill\cite{WHP03} use an analytical argument to predict that
$f_0$, the fraction of $E_{dw} = 0$ walls, is approximately 0.75,
independent of $p$, in the spin-glass regime.  However, the value of
$f_0$ depends strongly on the aspect ratio of the
lattice,\cite{HBCMY02,FH06} and it is not clear why this analytical
argument should apply only when the aspect ratio is equal to one.
It is also completely unclear to this author where the argument uses
the fact that $E_{dw} = 0$ domain walls can only occur when $L$ is
even.

Estimating the statistical uncertainties in the data precisely is
not trivial, due to the fact that the values of $E_{dw}$ obtained
from the same set of bonds with the four different pairs of boundary
conditions are not statistically independent.\cite{Note1}  An upper
bound on the statistical uncertainties is obtained by counting the
number of samples, rather than the number of McMillan pairs of
boundary conditions.
\begin{figure}
\includegraphics[width=3.1in]{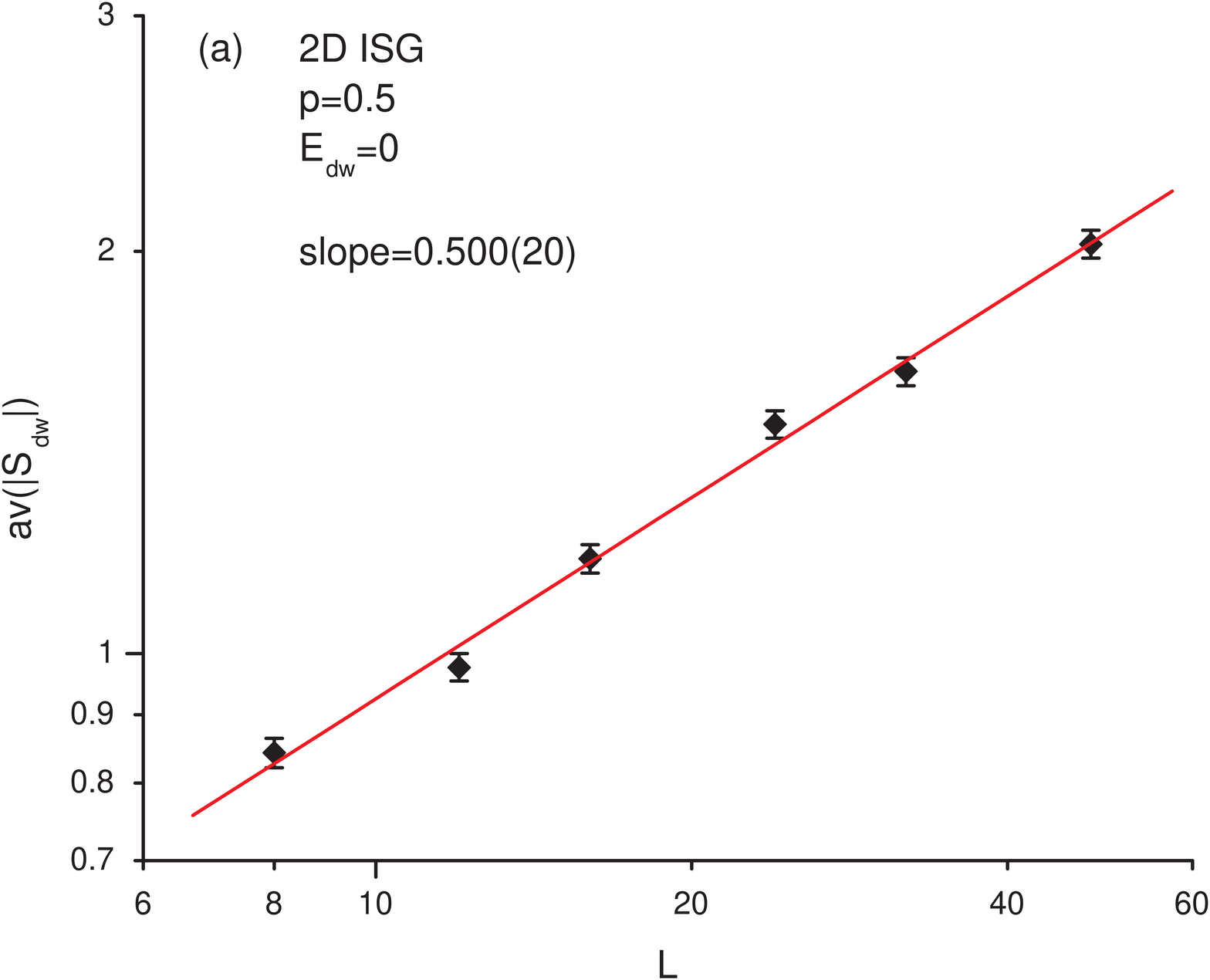}\quad
\includegraphics[width=3.1in]{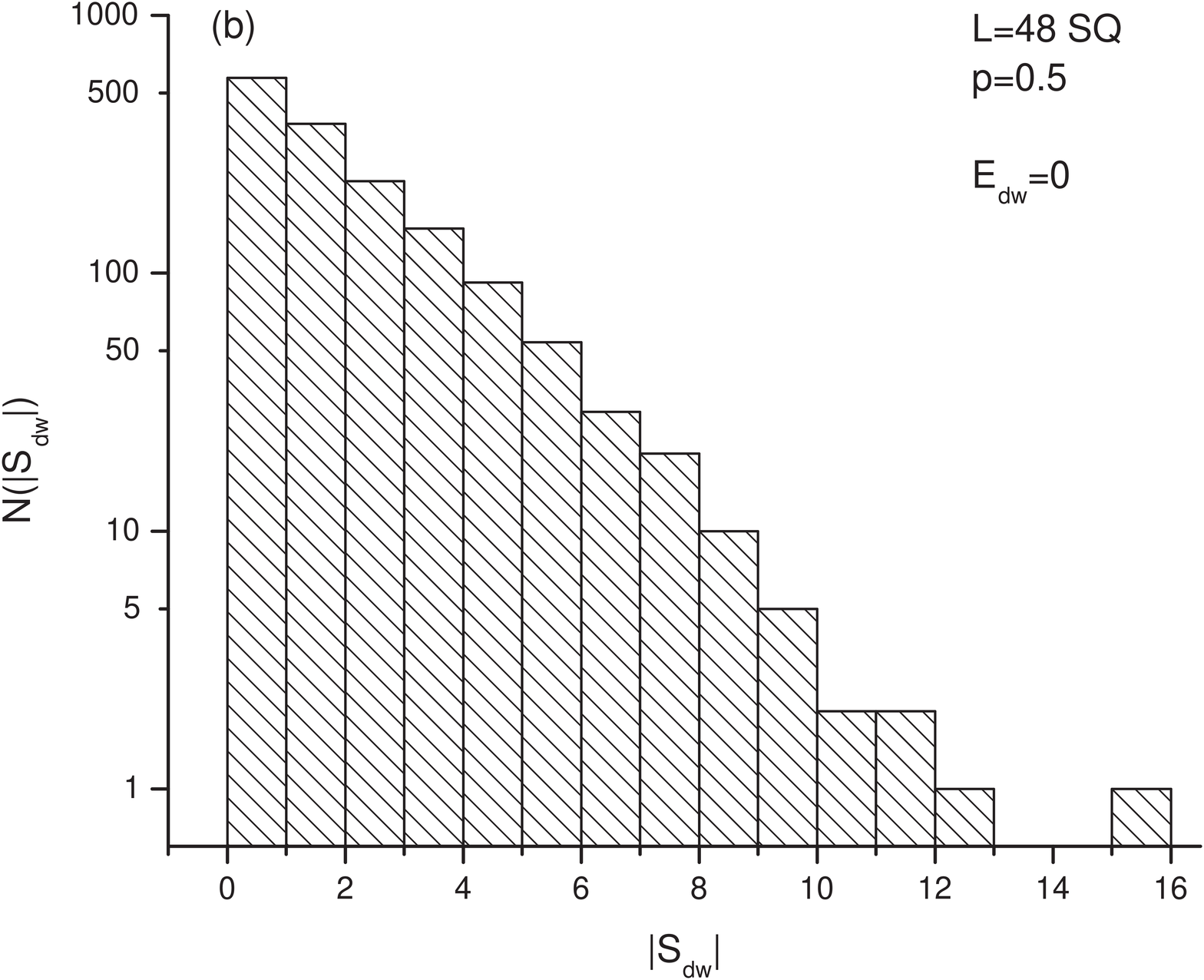}
\caption{\label{Fig.1}(color online) (a) Average $|S_{dw}|$ vs. $L$
for the $E_{dw} = 0$ domain walls, log-log plot. The error bars
indicate one standard deviation. (b) Histogram of $|S_{dw}|$ for
$E_{dw} = 0$ with $L = 48$.  The vertical scale is logarithmic.}
\end{figure}

The probability distribution of $S_{dw}$ for the cases where $E_{dw}
= 0$ should be symmetric about 0, and our statistics are consistent
with this.  If we look at the $L$-dependence of $[|S_{dw}|]$, shown
in Fig.~1(a), we find a scaling exponent
\begin{equation}
\theta_S ( 0 ) = 0.500 \pm 0.020  \,
\end{equation}
for $E_{dw} = 0$.  The result of Saul and Kardar,\cite{SK93,SK94}
obtained by looking at the distribution of $S_{dw}$ for all values
of $E_{dw}$ combined, was $\theta_S = 0.49 \pm 0.02$.  To obtain
this exponent, Saul and Kardar fit their data at small values of
$S_{dw}$.  When $L$ is even, which was the case for all of their
data, this part of the data belongs almost entirely to the $E_{dw} =
0$ component.\cite{Fis05}

The calculated means and skewness of these essentially symmetric
distributions for $S_{dw}$ is, naturally, consistent with zero, but
their kurtosis is not.  The reason for this is shown in Fig.~1(b),
which is a histogram for $|S_{dw}|$ of $E_{dw} = 0$ when $L = 48$.
We see that the distribution is approximately exponential, and
therefore far from Gaussian.  The computed kurtosis of this $L = 48$
distribution is 2.0, somewhat less than the value of 3 which would
be found for an exact two-sided exponential distribution.  The basic
shape of these distributions is similar for the smaller values of
$L$, with the width of each distribution given by the square root of
its variance.

\begin{figure}
\includegraphics[width=3.1in]{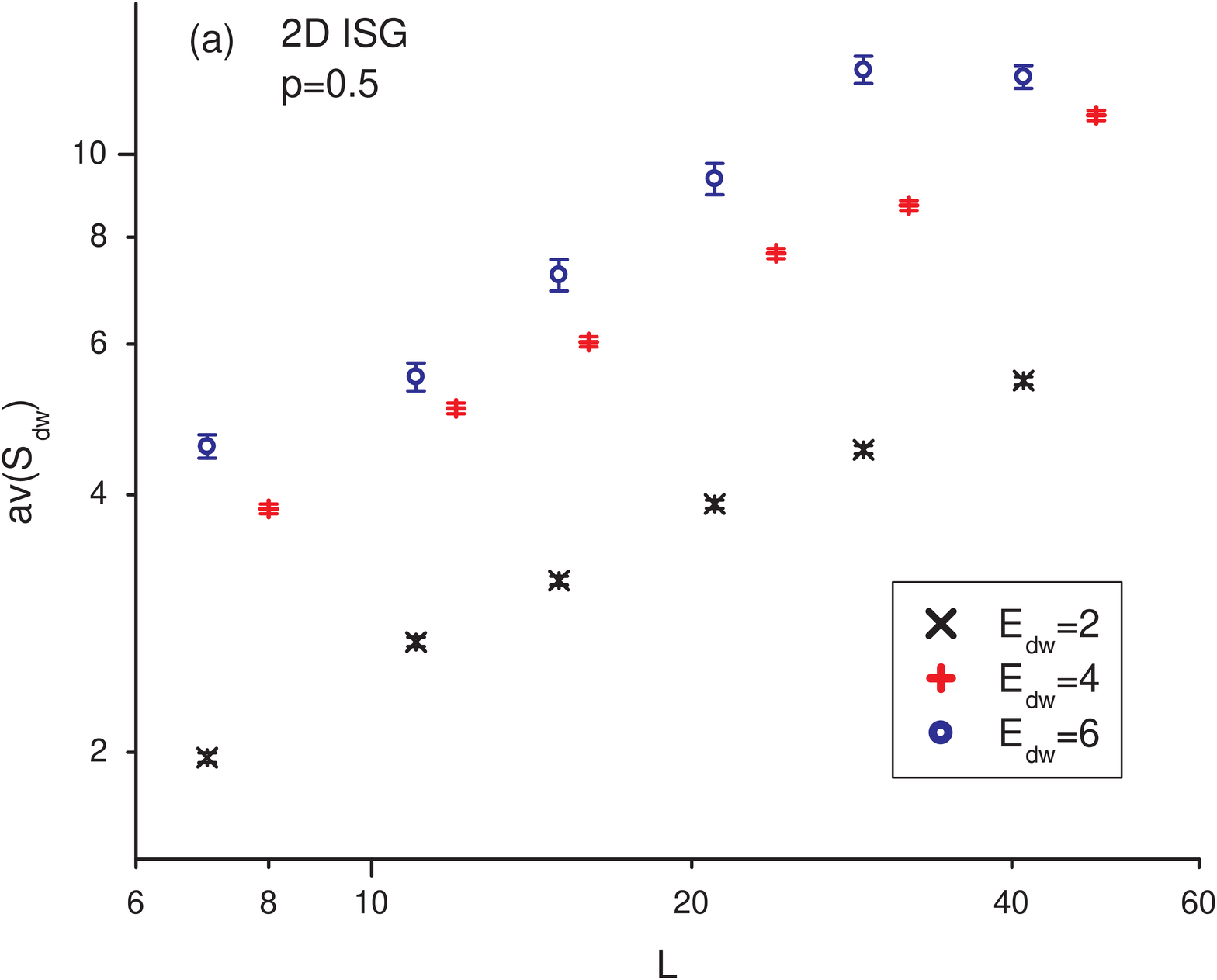}\quad
\includegraphics[width=3.1in]{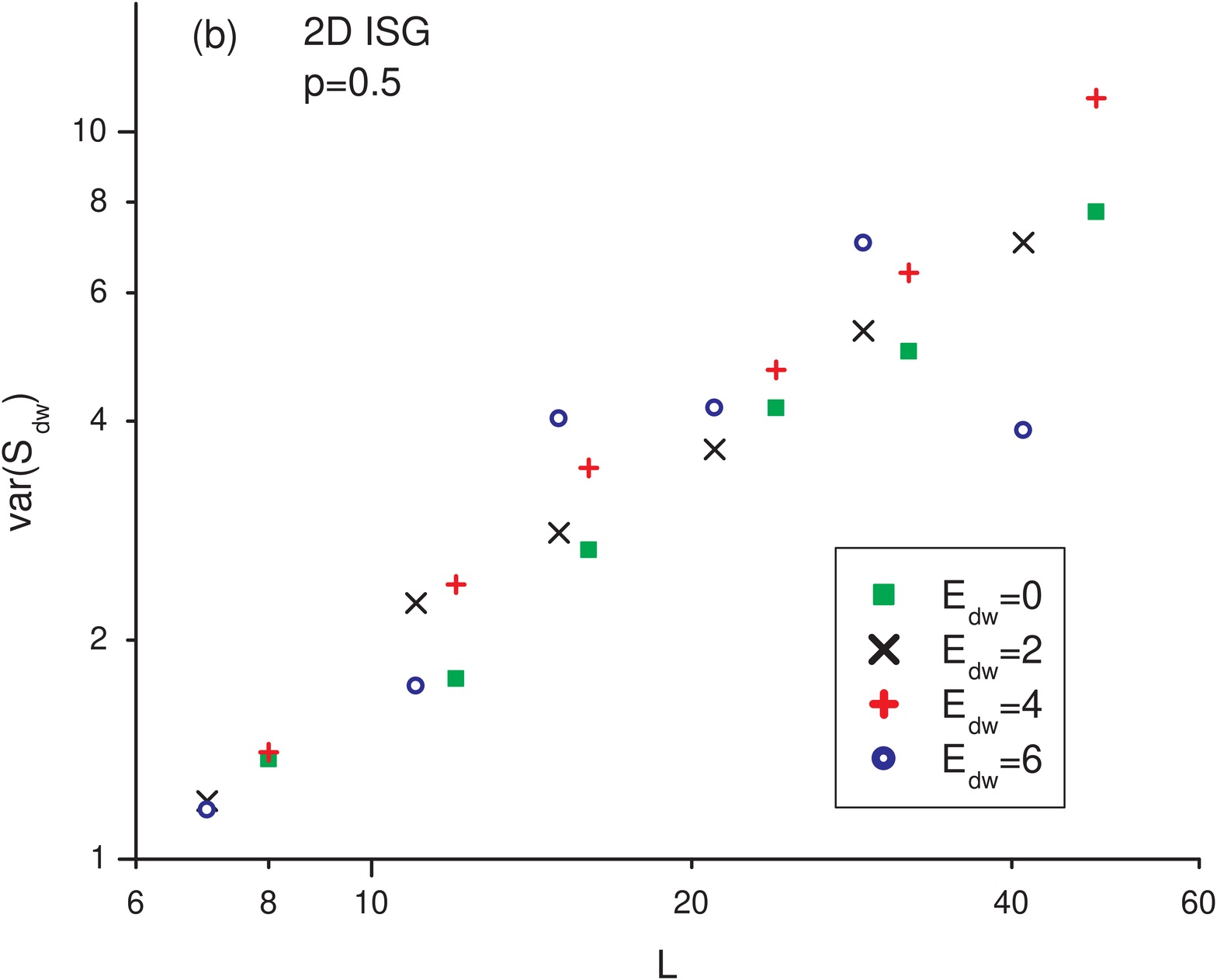}
\caption{\label{Fig.2}(color online) (a) Average $S_{dw}$ vs. $L$
for the $E_{dw}
> 0$ domain walls, log-log plot. The error bars indicate one
standard deviation. (b) Variance of $S_{dw}$ vs. $L$, log-log plot.}
\end{figure}

\begin{table}
\caption{\label{tab:table3} Scaling exponents for the first and
second cumulants of the $S_{dw}$ distributions.  $\theta_S$ is the
scaling exponent for $[S_{dw}]$, and $\phi_S$ is the scaling
exponent for the variance of ($S_{dw}$).}
\begin{ruledtabular}
\begin{tabular}{ccc}
 $E_{dw}$ & $\theta_S$ & $\phi_S$ \\
\hline
 0 & $0.500 \pm 0.020$ & $0.992 \pm 0.047$ \\
 2 & $0.565 \pm 0.019$ & $0.972 \pm 0.051$ \\
 4 & $0.584 \pm 0.015$ & $1.107 \pm 0.047$ \\
 6 & $0.617 \pm 0.062$ & $0.85 \pm 0.28$ \\

\end{tabular}
\end{ruledtabular}
\end{table}

When $E_{dw}$ is not zero, the relative signs of $E_{dw}$ and
$S_{dw}$ are not arbitrary.  Having chosen $E_{dw}$ to be
nonnegative, we then find that, when $E_{dw}$ is positive, it turns
out that $S_{dw}$ is usually positive.  In Fig.~2(a) we show the
behavior of $[S_{dw}]$ for the cases where $E_{dw} = k$, with $k =$
2, 4 and 6, as a function of $L$.  We see that for $k
> 0$, the average value of $S_{dw} ( L )$ grows approximately as
$L^{0.58}$. More precisely, least-squares fits to the form
\begin{equation}
[ S_{dw} ( L ) ] = A L^{\theta_S}  \,
\end{equation}
gives the results for $\theta_S ( k )$ shown in Table~III. The
result for $k = 6$ is rather uncertain, due to the small number of
examples of this type.  These results are consistent with the
prediction of droplet theory,\cite{FH88} that $\theta_S$ should be
independent of $k$ (aside from the $k = 0$ case, which is clearly
exceptional). However, there also appears to be a tendency for
$\theta_S ( k )$ to increase as $k$ increases.  Therefore, the
possibility that $\theta_S \to 5/8$ as $k \to \infty$, which would
be consistent with Eqn.~(1), cannot be excluded by these data.

Because of the large GS degeneracy in the $\pm J$ Ising spin glass,
one does not know how to compute $d_S$ directly for this model.
However, if we use the droplet model prediction, that there is a
single value for $\theta_S$, the result is not consistent with
Eqn.~(1). The author's opinion is that Eqn.~(1) must be generalized
when $\eta > 0$.

As shown by Saul and Kardar,\cite{SK93,SK94} the variance of
$S_{dw}$ when $E_{dw} = 0$ increases with $L$ in approximately a
linear fashion.  Calculating the variance of these distributions,
and using linear least squares fits on the log-log plot shown in
Fig.~2(b), we find that assuming the increase of the variance with
$L$ is a power law gives the results shown in Table~III. These
numbers are reasonably consistent with the hypothesis that the
scaling exponent for the variance of the $S_{dw}$ distributions is
equal to 1, independent of $E_{dw}$. It is also interesting to
observe that the magnitude of the variance, and not merely the slope
of the fit, seems to be independent of $E_{dw}$. Except in the
special $E_{dw} = 0$ case, $2 \theta_S$ is greater than 1.
Therefore, the exponent $d_S$ should be controlled by $\theta_S$, as
predicted by Eqn. (2).

\section{SCALING OF THE DISTRIBUTIONS}

\begin{figure}
\includegraphics[width=3.1in]{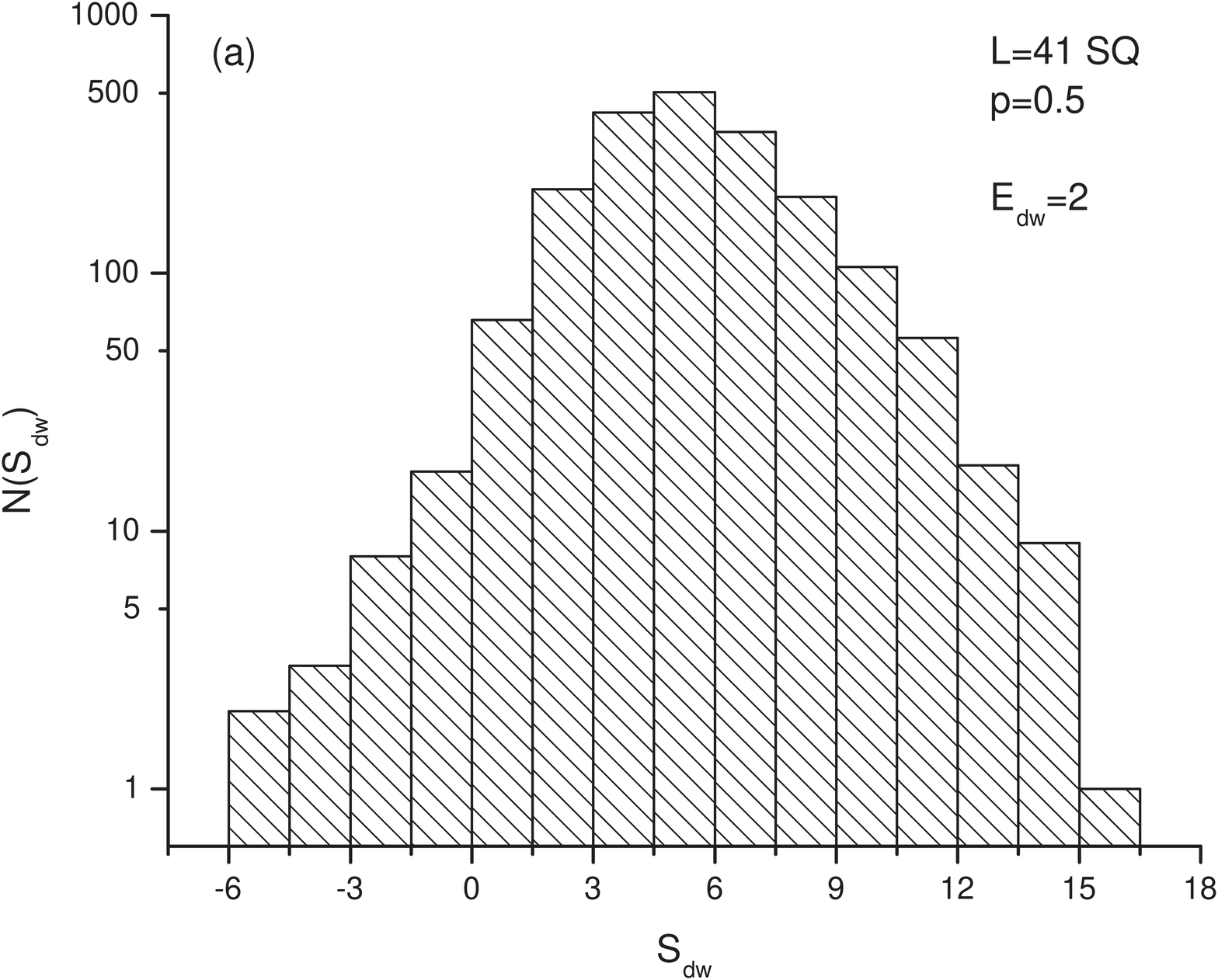}\quad
\includegraphics[width=3.1in]{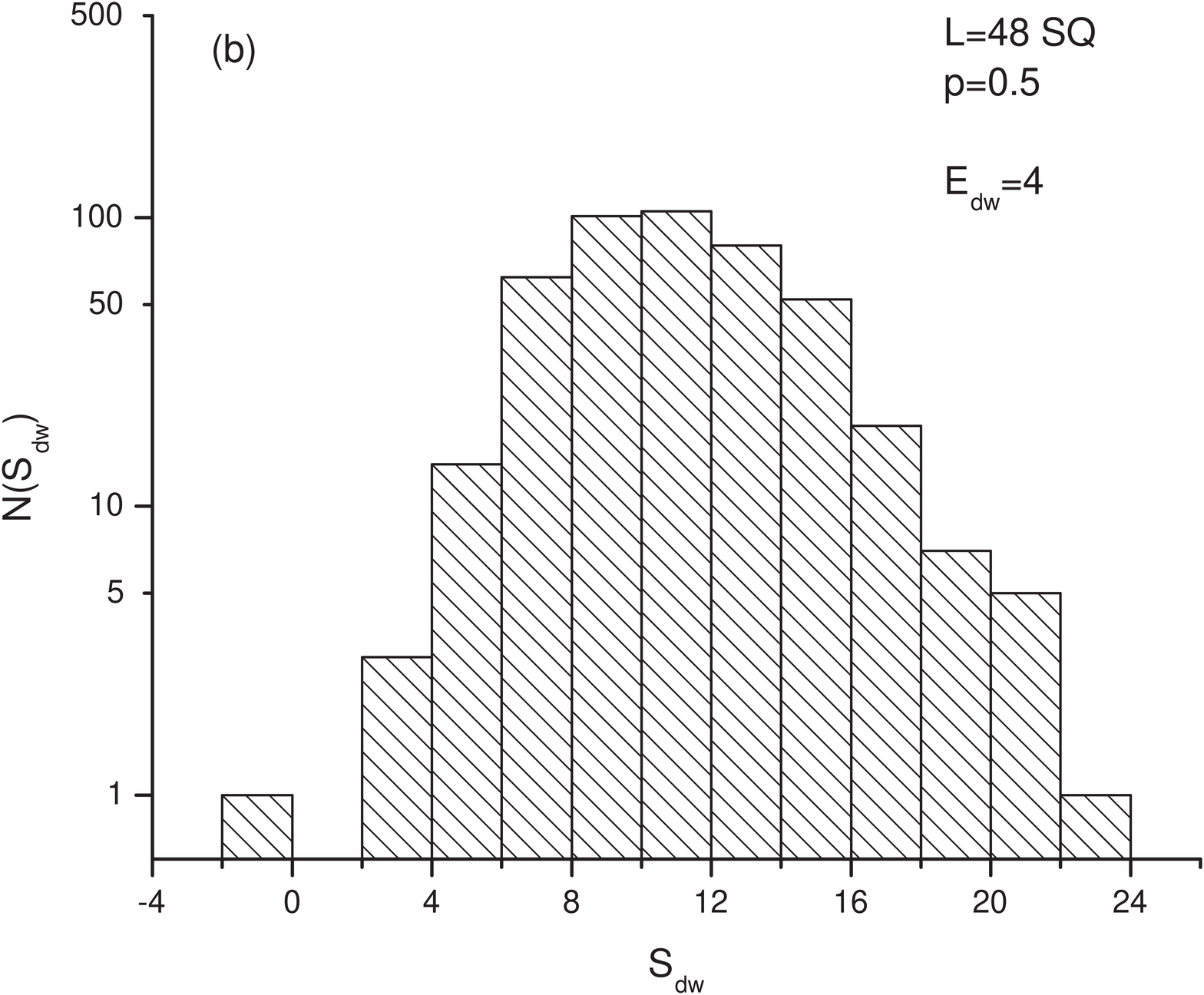}
\caption{\label{Fig.3} Histograms of $S_{dw}$ for (a) $E_{dw} = 2$
with $L = 41$ and (b) $E_{dw} = 4$ with $L = 48$.  The vertical
scales are logarithmic.}
\end{figure}

In Fig.~3 we show histograms for the $S_{dw}$ distributions for
$E_{dw} = 2$ at $L = 41$ and $E_{dw} = 4$ at $L = 48$.  In contrast
to the $E_{dw} = 0$ case, the skewness and kurtosis of the $S_{dw}$
distributions for $E_{dw} > 0$ are both small.  It is possible that
these distributions become Gaussian in the large $L$ limit. However,
the author is not aware of any reason why this must happen.

The basic shapes of the histograms in Fig.~3(a) and Fig.~3(b) appear
to be the same.  Since $2 \theta_S > 1$, it seems that the histogram
for the $E_{dw} = 4$ case can be mapped onto the histogram for the
$E_{dw} = 2$ case at a larger $L$.  A way of expressing this is that
for large $L$ the $S_{dw}$ histograms for $E_{dw} = k~>~0$ should
obey one-parameter scaling in the dimensionless variables
\begin{equation}
g_k ( L ) = {[S_{dw}]^2 \over [(S_{dw})^2]~-~[S_{dw}]^2} \, .
\end{equation}
If $\theta_S$ is independent of $k$, then, assuming $\phi_S = 1$,
\begin{equation}
g_k ( L ) = (L / L_k)^ {2 \theta_S - 1} \, ,
\end{equation}
where we define $L_k$ by the condition $g_k ( L_k ) = 1$.

What we have learned is that in this model there appear to be two
distinct classes of domain walls, the $E_{dw} = 0$ domain walls and
the $E_{dw} > 0$ domain walls.  As we have seen, the $E_{dw} > 0$
domain walls behave in a way which appears to be essentially
consistent with the predictions of the droplet model, but the
$E_{dw} = 0$ domain walls do not.  This difference in behavior is
due to the symmetry of the $E_{dw} = 0$ case, which forces the
average $S_{dw}$ to be zero.

For an $E_{dw} > 0$ domain wall, a large contribution to $S_{dw}$
comes from the shift in the average GS entropy with the shift in the
GS energy.\cite{Fis06}  What remains to be understood is why
$[S_{dw}]$ should scale with $L$ in the way predicted by the droplet
model.  The conventional derivation of the droplet model\cite{FH86}
uses the assumption that the GS is unique, up to a reversal of the
entire state, in an essential way.  What follows immediately from
this is that $\eta = 0$.  An extension of the droplet model to the
more general case was given by Fisher and Huse.\cite{FH88}  However,
the author hopes that by now he has convinced the reader that a
better understanding of the $\eta > 0$ case is needed.

\section{SUMMARY}

We have studied the statistics of domain walls for ground states of
the 2D Ising spin glass with +1 and -1 bonds for $L \times L$ square
lattices with $L \le 48$, and $p$ = 0.5, where $p$ is the fraction
of negative bonds, using periodic and/or antiperiodic boundary
conditions, for both even and odd $L$.  Under these conditions, most
domain walls have an energy $E_{dw} < 8$.  The probability
distribution of the entropy, $S_{dw}$, is found to depend strongly
on $E_{dw}$, but it appears possible to parameterize this dependence
in a simple way. The results for $S_{dw}$ do not appear to agree
quantitatively with the prediction of Amoruso, Hartmann, Hastings
and Moore,\cite{AHHM06} Eqn.~(1). Our results for $[|S_{dw}|]$ when
$E_{dw} = 0$ agree with those of Saul and Kardar,\cite{SK93,SK94}
but in addition we find that the distributions are close to being
exponential in that case, even in the limit of large $L$.  Due to
the special role of the $E_{dw} = 0$ domain walls, we can understand
the difference between the scaling exponent found by Saul and Kardar
and the prediction of the droplet model.

\begin{acknowledgments}
The author thanks J. Vondr\'{a}k for providing a copy of his
computer code, and for help in learning how to use it.  M. Kardar
and L. Saul provided unpublished details of their calculation. He is
grateful to S. L. Sondhi, A. K. Hartmann, D. F. M. Haldane, D. A.
Huse, J. Cardy and M. A. Moore, for helpful discussions, and to the
Physics Department of Princeton University for providing use of the
computers on which the data were obtained.

\end{acknowledgments}



\end{document}